\documentclass[pra,superscriptaddress,showpacs,twocolumn,10pt]{revtex4}
\usepackage{amssymb}
\usepackage{amsmath}
\usepackage{graphicx}

\begin{document}%

\newcommand{\ket}[1]{|#1\rangle}
\newcommand{\bra}[1]{\langle#1|}
\newcommand{\inner}[2]{\langle#1|#2\rangle}
\newcommand{\lr}\longrightarrow
\newcommand{\ra}\rightarrow
\newcommand{\tr}{{\rm Tr}}
\newcommand{\sgn}{{\rm sgn}}
\newcommand{\fsp}{{\rm span}}
\newcommand{\fsup}{{\rm supp}}
\newcommand{\fdg}{{\rm diag}}

\newtheorem{thm}{Theorem}
\newtheorem{Prop}{Proposition}
\newtheorem{Coro}{Corollary}
\newtheorem{Lemma}{Lemma}
\newtheorem{Def}{Definition}

\title{Universal programmable devices for unambiguous discrimination}
\author{Chi Zhang}
\email{zangcy00@mails.tsinghua.edu.cn}
\author{Mingsheng Ying}
\email{yingmsh@tsinghua.edu.cn}
\affiliation{State Key Laboratory
of Intelligent Technology and Systems, Department of Computer
Science and Technology Tsinghua University, Beijing, China,
100084}
\date{\today}

\begin{abstract}
We discuss the problem of designing unambiguous programmable
discriminators for any $n$ unknown quantum states in an
$m$-dimensional Hilbert space. The discriminator is a fixed
measurement which has two kinds of input registers: the program
registers and the data register. The program registers consist of
the $n$ states, while the data register is prepared among them.
The task of the discriminator is to tell us which state stored in
the program registers is equivalent to that in the data register.
First, we give a necessary and sufficient condition for judging an
unambiguous programmable discriminator. Then, if $m=n$, we present
an optimal unambiguous programmable discriminator for them, in the
sense of maximizing the worst-case probability of success.
Finally, we propose a universal unambiguous programmable
discriminator for arbitrary $n$ quantum states. We also show how
to use this universal discriminator to unambiguously discriminate
mixed states.
\end{abstract}

\maketitle

\section{Introduction}
Discrimination between quantum states is an essential task in
quantum communication protocols. Generally, a set of states cannot
be discriminated exactly, unless they are orthogonal to each other
\cite{book1}. One strategy of discriminating non-orthogonal
quantum states is the so-called unambiguous discrimination: with a
non-zero possibility of getting inconclusive answer, one can
distinguish the given states without error
\cite{IV,DI,PE,JS,CH,SZ,Eldar,Minimax,comparison}. Such a strategy
works if and only if the states to be distinguished are linearly
independent \cite{CH}, and finding the optimal unambiguous
discrimination through Bayesian approach with a given priori
probability distribution, can be reduced to a semi-definite
programming (SDP) problem \cite{SZ,Eldar}. On the other hand,
D'Ariano \textit{et al} \cite{Minimax} considered the problem of
finding optimal unambiguous discrimination through
\textquotedblleft minimax strategy''. In such a strategy, no
information about priori probability is given, and the
discriminator is designed to maximize the smallest of the success
probabilities.

All above discriminators depend on the set of states being
discriminated. When states change, the device also need to be
changed. Recently, the problem of designing programmable
discriminator attracted a lot of attention. In a programmable
quantum device, quantum states are input through two kinds of
registers: program registers and data registers. The states in
data registers are manipulated by the fixed device, according to
the states in program registers \cite{P1,P2,P3,P4,P5,P6,P7}.
Particularly, in a programmable discriminator, the information
about states being discriminated is offered through a
\textquotedblleft quantum program'', according to which, the
discrimination on the state in data register is specified.
Different from the discriminators for known states, a programmable
discriminator is capable to discriminate any states, with the
corresponding program. In Ref. \cite{program1}, Du\v{s}ek
\textit{et al} provided a model of unambiguous programmable
discriminator for a pair of $1$-qubit states. In this model, a new
quantum state, besides the pair of states being discriminated, is
needed for programming. Recently, Bergou \textit{et al}
\cite{Programmable} constructed an alternative unambiguous
programmable discriminator for any two different states. The
advantage of this discriminator is that, the \textquotedblleft
quantum program'' is simply comprised of the states being
discriminated. Furthermore, unambiguous programmable discriminator
for two states with a certain number of copies is also discussed
\cite{copy,copy2}. All of above tasks focus on discriminating two
states, and estimates the efficiency with a given priori
probability. In addition, Fiur\'{a}\v{s}ek \textit{et al}
\cite{JF} considered several kinds of programmable quantum
measurement devices, including a device performing von Neumann
measurement on a qudit, which can also be regarded as a
discriminator for orthogonal states.

In this paper, we describe the more general unambiguous
programmable discriminators for any $n$ quantum states. The
quantum program used in these discriminators is the tensor product
of the $n$ states being discriminated, so that there is no extra
states needed for programming. We design the optimal
discriminators in a minimax strategy, so that the optimal
discriminators are not dependent on any priori information. Since
quantum states can be unambiguously discriminated if and only if
they are linearly independent, we strict our discussion under this
condition, and claim a programmable discriminator
\textquotedblleft universal'' if it can unambiguously discriminate
any set of linearly independent states.

Our present article is organized as follows. Section.~\ref{1} is a
preliminary section in which we recall some results needed in the
sequel from linear algebra \cite{book}. In section.~\ref{2}, we
give a necessary and sufficient condition for unambiguous
programmable discriminators. Further, in section.~\ref{3}, we
define the efficiency of a discriminator under the minimax
strategy, and provide a set of properties for the optimal
discriminators. Then, we present the optimal unambiguous
programmable discriminators for $n$ arbitrary quantum states in an
$n$-dimensional Hilbert space in section.~\ref{4}, and propose a
set of unambiguous programmable discriminators for $n$ quantum
states in an $m$-dimensional Hilbert space, where $m>n$, in
section.~\ref{5}. Finally, we show how to utilize our scheme to
unambiguously discriminate mixed states, in section.~\ref{6}.
Section.~\ref{7} is a short summary.

\section{Preliminaries}\label{1}

Let us begin with some preliminaries that are useful in presenting
our main results.

The antisymmetric tensor product of states $\ket{\varphi_1}$,
$\ket{\varphi_2}$, $\cdots$, $\ket{\varphi_n}$ in a Hilbert space
$H$ is defined as
\begin{equation}
\begin{split}
&\ket{\varphi_1}\wedge\ket{\varphi_2}\wedge\cdots\wedge\ket{\varphi_n}\\
= &\frac{1}{\sqrt{n!}}\sum_{\sigma\in
S(n)}\sgn(\sigma)\ket{\varphi_{\sigma_1}}\ket{\varphi_{\sigma_2}}\cdots\ket{\varphi_{\sigma_n}},
\end{split}
\end{equation}
where $S(n)$ is the symmetric (or permutation) group of degree
$n$, $\sgn(\sigma)$ denotes the signature of permutation $\sigma$,
i.e., $\sgn(\sigma) = +1$, if $\sigma$ is an even permutation;
$\sgn(\sigma) = -1$, if $\sigma$ is an odd permutation. The span
of all antisymmetric tensors
$\ket{\varphi_1}\wedge\ket{\varphi_2}\wedge\cdots\wedge\ket{\varphi_n}$
in $H^{\otimes n}$ is denoted by $\wedge^n H$, which is called the
antisymmetric subspace of $H^{\otimes n}$. If the dimension of $H$
is $m$, then the dimension of $\wedge^n H$ is $C_m^n$.

In an $n$-composite system, for any $\sigma\in S(n)$, we also use
$\sigma$ to represent a linear operation on the system, which
realigns the subsystems according to $\sigma$, i.e.,
\begin{equation}
\sigma \ket{\varphi_1}\ket{\varphi_2}\cdots\ket{\varphi_n} =
\ket{\varphi_{\sigma_1}}\ket{\varphi_{\sigma_2}}\cdots\ket{\varphi_{\sigma_n}},
\end{equation}
here
$\ket{\varphi}=\ket{\varphi_1}\ket{\varphi_2}\cdots\ket{\varphi_n}$
is an arbitrary product state in the $n$-composite system. It is
easy to prove that $\sigma$ is a unitary operation. For a state
$\ket{\psi}\in H^{\otimes n}$, $\ket{\psi}\in\wedge^n H$ if and only
if for any $\sigma\in S(n)$,
\begin{equation}
\sigma\ket{\psi}=\sgn(\sigma)\ket{\psi}.
\end{equation}

In this paper, we denote the projector of $\wedge^n H$ by
$\Phi(n)$. For any product state
$\ket{\varphi}=\ket{\varphi_1}\ket{\varphi_2}\cdots\ket{\varphi_n}$
in $H^{\otimes n}$,
\begin{equation}\label{pro}
\bra{\varphi}\Phi(n)\ket{\varphi} = \frac{1}{n!}\det(X),
\end{equation}
where $X$ is the Gram matrix of
$\{\ket{\varphi_1},\ket{\varphi_2},\cdots,\ket{\varphi_n}\}$, i.e.,
the $(i,j)$ element of $X$ is
\begin{equation}
X_{(i,j)} = \inner{\varphi_i}{\varphi_j}.
\end{equation}
Hence, the Eq.(\ref{pro}) equals to zero if and only if
$\{\ket{\varphi_1},\ket{\varphi_2},\cdots,\ket{\varphi_n}\}$ are
linearly dependent.

\section{Unambiguous Programmable Discriminator}\label{2}
An unambiguous programmable discriminator for $n$ quantum states in
an $m$-dimensional Hilbert space $H$, can be simply designed in the
following version. The discriminator has $n$ program registers and
one data register. When the quantum state wanted to be identified is
selected in states $\ket{\psi_1},\ldots,\ket{\psi_n}$, the $i$th
program register is put in the state $\ket{\psi_i}$, for
$i=1,\ldots,n$, and the data register is prepared in the state
wanted to be identified.  Here, we label the $i$th program register
as the $i$th subsystem, the data register as the $(n+1)$th
subsystem, and use $\bar{i}$ to indicate the system consisting of
all subsystems under consideration except the $i$th one. For
simplicity, we introduce a notation $\ket{\alpha^s_t}$ to denote a
special kind of product states in a $(s+1)$-component quantum
system, where the state in the $l$th subsystem is $\ket{\alpha_l}$,
for any $1\leq l\leq s$, and the state in the $s+1$ subsystem is the
same as the $t$th subsystem, i.e.,
\begin{equation}
\ket{\alpha^s_t}=\ket{\alpha_1}\ket{\alpha_2}\cdots\ket{\alpha_s}\ket{\alpha_t}.
\end{equation}
Then, if the data register is in $\ket{\psi_j}$, the total input
state is $
\ket{\psi^n_j}=\ket{\psi_1}\ket{\psi_2}\cdots\ket{\psi_n}\ket{\psi_j}$.
The discriminator is described by a general POVM
$\{\Pi_0,\Pi_1,\cdots,\Pi_n\}$ on the entire input system,
including all program registers and the data register. For any $i
\neq j$, $i \neq 0$, if it is satisfied that
$\bra{\psi_j^n}\Pi_i\ket{\psi_j^n} = 0$, then when outcome $i$ $(i
\neq 0)$ is observed, one may claim with certainty that the data
register is originally prepared in the state $\ket{\psi_i}$, and
occurrence of outcome 0 means that the identification fails to
give a report. In this paper, we also use $\overrightarrow{\Pi}$
to denote the measurement $\{\Pi_0,\Pi_1,\cdots,\Pi_n\}$ for
simplicity.

The main purpose of this section is to present a necessary and
sufficient condition for unambiguous programmable discriminators. We
would like to start with a lemma for positive operators, which will
be used in the proof for the condition.

\begin{Lemma}\label{uneq}
Suppose $\Omega$ is a positive operator on a composite system AB,
for any product state $\ket{\varphi} =
\ket{\varphi_a}_A\ket{\varphi_b}_B$, it holds that
\begin{equation}
\bra{\varphi}\Omega\ket{\varphi}\tr(\Omega) \leq
\bra{\varphi_a}\tr_B(\Omega)\ket{\varphi_a}\bra{\varphi_b}\tr_A(\Omega)\ket{\varphi_b}.
\end{equation}
\end{Lemma}
{\it Proof.} It is observed that $\Omega/\tr(\Omega)$ satisfies
the trace condition and positivity condition for a density
operator. Let $\rho = \Omega/\tr(\Omega)$, which is a density
operator, and consider a quantum operation $\varepsilon = \tr_B
\otimes \tr_A$. Then
\begin{equation}
F(\varepsilon(\rho),\varepsilon(\ket{\varphi}\bra{\varphi})) \leq
F(\rho,\ket{\varphi}\bra{\varphi}),
\end{equation}
where $F$ stands for the fidelity between two density operators
\cite{book1}. Because $\ket{\varphi}$ is a pure product state, we
have that
\begin{equation}
\begin{split}
&\bra{\varphi}\Omega\ket{\varphi}\tr(\Omega)\\
= &\bra{\varphi}\rho\ket{\varphi}(\tr(\Omega))^2\\ \leq
&\bra{\varphi_a}\tr_B(\rho)\ket{\varphi_b}\bra{\varphi_b}\tr_A(\rho)\ket{\varphi_b}(\tr(\Omega))^2\\
=
&\bra{\varphi_a}\tr_B(\Omega)\ket{\varphi_a}\bra{\varphi_b}\tr_A(\Omega)\ket{\varphi_b}.
\end{split}
\end{equation}
This completes the proof. \hfill $\Box$

\begin{thm}\label{struct}
A measurement $\{\Pi_0,\Pi_1,\cdots,\Pi_n\}$ is an unambiguous
programmable discriminator for any $n$ quantum states in Hilbert
space $H$, if and only if the support space of $\tr_i(\Pi_i)$ is a
subspace of $\wedge^n H$, i.e.,
\begin{equation}\label{stru1}
\fsup(\tr_i(\Pi_i)) \leq {\wedge}^n H,
\end{equation}
where $\tr_i$ is the partial trace over the $i$th subsystem, and
$\wedge^n H$ is the antisymmetric subspace of $H^{\otimes n}$.
\end{thm}

{\it Proof.} ``$\Longrightarrow$". Suppose $\ket{\phi}$ is an
arbitrary eigenvector of $\Pi_i$ with non-zero eigenvalue, since
$\Pi_i$ is a positive operator,
\begin{equation}\label{15}
\inner{\psi^n_j}{\phi} = 0,
\end{equation}
for any $j\neq i$. To prove Eq.(\ref{stru1}), we only have to
prove that
\begin{equation}\label{wat}
\fsup(\tr_i(\ket{\phi}\bra{\phi})) \leq {\wedge}^n H.
\end{equation}

Let $\{\ket{1},\ket{2},\ldots,\ket{m}\}$ be an orthonormal basis
for Hilbert space $H$. As $\ket{\phi}\in H^{\otimes(n+1)}$, it can
be rewritten as
\begin{equation}
\ket{\phi} = \sum_{\omega} \upsilon(\omega)\ket{\omega},
\end{equation}
where $\ket{\omega}$ is the orthonormal basis of space
$H^{\otimes(n+1)}$, derived from the given basis of $H$, i.e.,
\begin{equation}
\ket{\omega} =
\ket{\omega_1}\ket{\omega_2}\cdots\ket{\omega_{n+1}},
\end{equation}
where $\ket{\omega_k}\in \{\ket{1},\ket{2},\ldots,\ket{m}\}$,
$1\leq k\leq n$, and $v(\omega)$ is the corresponding coefficient.
Because Eq.(\ref{15}) should be satisfied with any input states
under consideration, we can choose some special states to derive
necessary conditions for $\ket{\phi}$.

First, choose $\ket{\psi_j}=\ket{s}$, where $\ket{s}\in
\{\ket{1},\cdots,\ket{m}\}$,
\begin{equation}\label{pr1}
\begin{split}
\inner{\psi^n_j}{\phi} &=
\sum_{\omega}v(\omega)\inner{s}{\omega_j}\inner{s}{\omega_{n+1}}\prod_{k\neq
j}\inner{\psi_k}{\omega_k}\\
&= \sum_{\omega_{\bar{j}}}
v(\omega|\omega_j=\omega_{n+1}=s,\omega_{\bar{j}})\inner{\psi_{\bar{j}}}{\omega_{\bar{j}}},
\end{split}
\end{equation}
where
\begin{equation}
\ket{\psi_{\bar{j}}} =
\ket{\psi_1}\ket{\psi_2}\cdots\ket{\psi_{j-1}}\ket{\psi_{j+1}}\cdots\ket{\psi_n},
\end{equation}
and
\begin{equation}
\ket{\omega_{\bar{j}}} =
\ket{\omega_1}\ket{\omega_2}\cdots\ket{\omega_{j-1}}\ket{\omega_{j+1}}\cdots\ket{\omega_n}.
\end{equation}
Since $\ket{\psi_{\bar{j}}}$ can be any product state in $H^{\otimes
(n-1)}$ and $\ket{\omega_{\bar{j}}}$s form an orthonormal basis for
$H^{\otimes (n-1)}$, to confirm that Eq.(\ref{pr1}) always equals to
zero, it must holds that, $v(\omega) = 0$, if $\omega_j =
\omega_{n+1}$, for some $j\neq i$.

Next, choose $\ket{\psi_j}=\frac{1}{\sqrt{2}}(\ket{s}+\ket{t})$,
where $\ket{s}, \ket{t}\in \{\ket{1},\ket{2},\ldots,\ket{m}\}$,
\begin{equation}\label{pr2}
\begin{split}
\inner{\psi^n_j}{\phi}
=\sum_{\omega}&v(\omega)\inner{\psi_j}{\omega_j}\inner{\psi_{n+1}}{\omega_{n+1}}\prod_{k\neq
j}\inner{\psi_k}{\omega_k}\\
=
\frac{1}{2}\sum_{\omega_{\bar{j}}}\big{(}&v(\omega|\omega_j=\omega_{n+1}=s,\omega_{\bar{j}})\\
+&v(\omega|\omega_j=\omega_{n+1}=t,\omega_{\bar{j}})\\
+&v(\omega|\omega_j=s,\omega_{n+1}=t,\omega_{\bar{j}})\\
+&v(\omega|\omega_j=t,\omega_{n+1}=s,\omega_{\bar{j}})\big{)}\inner{\psi_{\bar{j}}}{\omega_{\bar{j}}}\\
=\frac{1}{2}\sum_{\omega_{\bar{j}}}\big{(}&v(\omega|\omega_j=s,\omega_{n+1}=t,\omega_{\bar{j}})\\
+&v(\omega|\omega_j=t,\omega_{n+1}=s,\omega_{\bar{j}})\big{)}\inner{\psi_{\bar{j}}}{\omega_{\bar{j}}},
\end{split}
\end{equation}
where $\ket{\psi_{\bar{j}}},\ket{\omega_{\bar{j}}}$ have the same
meanings as those in Eq.(\ref{pr1}). Therefore, we have that
$v(\omega)+v((j,n+1)\omega)=0$, for any $j\neq i$, where
$(j,n+1)\omega$ represents the sequence obtained by exchanging the
$j$th and the $(n+1)$th elements in $\omega=\omega_1 \omega_2 \ldots
\omega_{n+1}$. Because $(j,k) = (j,n+1)(k,n+1)(j,n+1)$, it is
derived that
\begin{equation}\label{anti}
v(\omega) + v((j,k)\omega) = 0,
\end{equation}
for any $j,k$ different from $i$.

To proceed, we partition the total input system into two subsystems,
the first one is the $i$th program register, and the second one
include the rest $n-1$ program registers and the data register. We
use $i$ and $\bar{i}$ to denote these subsystems respectively, then
\begin{equation}\label{phi}
\begin{split}
\ket{\phi} &= \sum_{\omega}
v(\omega)\ket{\omega_i}_i\ket{\omega'}_{\bar{i}}\\
&=
\sum_{s=1}^{m}\ket{s}_i\sum_{\omega'}v(\omega|\omega_i=s,\omega')\ket{\omega'}_{\bar{i}}\\
&= \sum_{s=1}^{m} \ket{s}_i\ket{\phi'_s}_{\bar{i}},
\end{split}
\end{equation}
where
\begin{equation}
\ket{\omega'}=\ket{\omega_1}\ket{\omega_2}\cdots\ket{\omega_{i-1}}\ket{\omega_{i+1}}\cdots\ket{\omega_n}\ket{\omega_{n+1}},
\end{equation}
and
\begin{equation}\label{phi'}
\ket{\phi'_s} =
\sum_{\omega'}v(\omega|\omega_i=s,\omega')\ket{\omega'}.
\end{equation}

The support space of $\tr_i(\ket{\phi}\bra{\phi})$ is the span
space of $\ket{\phi'_s}$, for $1\leq s\leq m$. From
Eq.(\ref{anti}),
\begin{equation}\label{e1}
\inner{\omega'}{\phi'_s}= -\inner{(j,k)\omega'}{\phi'_s},
\end{equation}
where $j\neq k$. Hence, for any $\sigma\in S(n)$,
\begin{equation}
\sigma\ket{\phi'_s} = \sgn(\sigma)\ket{\phi'_s}.
\end{equation}
which means that $\ket{\phi'_s}$ is in $\wedge^n H$, for any
$1\leq s\leq m$. Then Eq.(\ref{wat}) is satisfied, and the support
space of $\tr_i(\Pi_i)$ is in $\wedge^n H$.

``$\Longrightarrow$". We also divide the total input system into
two subsystems: the $i$th program register labeled by $i$, and the
rest program registers and the data register labeled by $\bar{i}$.
When $j\neq i$, the total input state
\begin{equation}
\ket{\psi^n_j} = \ket{\psi_i}_i\ket{\psi'}_{\bar{i}},
\end{equation}
where
\begin{equation}
\ket{\psi'} =
\ket{\psi_1}\ket{\psi_2}\cdots\ket{\psi_{i-1}}\ket{\psi_{i+1}}\cdots\ket{\psi_n}\ket{\psi_j}.
\end{equation}
Because there are two $\ket{\psi_j}$s in the sequence
$\ket{\psi_1}$,$\ket{\psi_2}$,$\cdots$,
$\ket{\psi_{i-1}}$,$\ket{\psi_{i+1}}$,$\cdots$,$\ket{\psi_n}$,$\ket{\psi_j}$,
the states in this sequence are linearly dependent, from
Eq.(\ref{pro}),
\begin{equation}\label{tt}
\bra{\psi'}\Phi(n)\ket{\psi'} = 0.
\end{equation}
From Lemma.\ref{uneq},
\begin{equation}
\bra{\psi^n_j}\Pi_i\ket{\psi^n_j} \leq
\bra{\psi_i}\tr_{\bar{i}}(\Pi_i)\ket{\psi_i}\bra{\psi'}\tr_i(\Pi_i)\ket{\psi'}/\tr(\Pi_i).
\end{equation}
Since $\tr_i(\Pi_i)\leq\wedge^n H$, from Eq.(\ref{tt}),
\begin{equation}
\bra{\psi^n_j}\Pi_i\ket{\psi^n_j} = 0,
\end{equation}
for any $j\neq i$, note that $\Pi_i$ is a positive operator.
Therefore, $\{\Pi_0,\Pi_i,\cdots,\Pi_n\}$ can unambiguously
discriminate an arbitrary set of states
$\{\ket{\psi_1},\ket{\psi_2},\cdots,\ket{\psi_n}\}$, by the
quantum program $\ket{\psi_1}\ket{\psi_2}\cdots\ket{\psi_n}$.
\hfill $\Box$

The term ``unambiguous'' used here is in a generalized sense. When
a discriminator is claimed to be unambiguous, it only means that
the discriminator never make an error, however, it may always give
an inconclusive answer. For example, when $m>n$, consider a
measurement $\{\Pi_0,\cdots,\Pi_n\}$, such that for any $i\neq 0$,
$\Pi_i = \frac{1}{n}\Phi(n+1)$, where $\Phi(n+1)$ is the projector
of $\wedge^{(n+1)} H$. In this measurement
\begin{equation}
\bra{\psi^n_j}\Pi_i\ket{\psi^n_j} = 0,
\end{equation}
for any $i,j$, where $i\neq 0$. Hence, it is an unambiguous
programmable discriminator, however, the success probability of
identifying the state is always zero.

\section{minimax strategy for designing optimal
discriminator}\label{3}

Note that when a programmable discriminator is designed, no
information about states which would be discriminated by this
device is given. Thus, it is reasonable to find the optimal
discriminator in a minimax approach. In this strategy, the
optimal discriminator is designed to maximize the minimum success
probability of discriminating one state from an arbitrary set.
For a given measurement, the discrimination efficiency would be
defined as
\begin{equation}\label{def}
p(\overrightarrow{\Pi}) = \min_{\{\ket{\psi_i}\}}\min_i
p_i(\overrightarrow{\Pi}),
\end{equation}
where $\overrightarrow{\Pi}$ is the measurement satisfying the
condition for unambiguous programmable discriminator,
$\{\ket{\psi_i}\}$ ranges over all state sets that are linearly
independent, and $p_i(\overrightarrow{\Pi})$ is the success
probability of identifying the $i$th state $\ket{\psi_i}$, by the
measurement $\overrightarrow{\Pi}$, i.e.,
\begin{equation}
p_i(\overrightarrow{\Pi}) = \bra{\psi^n_i}\Pi_i\ket{\psi^n_i}.
\end{equation}

It is observed that the unambiguous programmable discriminators
for $n$ quantum states form a convex set. For any
$0\leq\lambda\leq 1$, if $\overrightarrow{\Pi}$ and
$\overrightarrow{\Pi'}$ are two POVM satisfying the conditions for
unambiguous programmable discriminators, $\overrightarrow{\Xi} =
\lambda\overrightarrow{\Pi}+(1-\lambda)\overrightarrow{\Pi'}$ is
also an unambiguous programmable discriminator. Furthermore, the
success probability of identifying the $i$th state in a given
state set by $\overrightarrow{\Xi}$,
\begin{equation}
\begin{split}
p_i(\overrightarrow{\Xi}) &=
\bra{\psi^n_i}(\lambda\Pi_i+(1-\lambda)\Pi'_i)\ket{\psi^n_i}\\
&=
\lambda\bra{\psi^n_i}\Pi_i\ket{\psi^n_i}+(1-\lambda)\bra{\psi^n_i}\Pi'_i\ket{\psi^n_i}\\
&= \lambda
p_i(\overrightarrow{\Pi})+(1-\lambda)p_i(\overrightarrow{\Pi'}),
\end{split}
\end{equation}
which is the corresponding convex combination of the success
probabilities of identifying the state by $\overrightarrow{\Pi}$
and $\overrightarrow{\Pi'}$. Then,
\begin{equation}\label{ueq}
\begin{split}
p(\overrightarrow{\Xi}) &= \min_{\{\ket{\psi_i}\}}\min_i
p_i(\overrightarrow{\Xi})\\
&= \min_{\{\ket{\psi_i}\}}\min_i \lambda
p_i(\overrightarrow{\Pi})+(1-\lambda)p_i(\overrightarrow{\Pi'})\\
&\geq \lambda p(\overrightarrow{\Pi}) +
(1-\lambda)p(\overrightarrow{\Pi'}).
\end{split}
\end{equation}
Hence, the efficiency of the unambiguous programmable
discriminator is a concave function.

In the remainder of this section, we provide some properties for
optimal unambiguous programmable discriminators.

\begin{Lemma}\label{LU}
Suppose $\overrightarrow{\Pi}$ is the optimal unambiguous
programmable discriminator for $n$ states in Hilbert space $H$,
then for any unitary operator $U$ in $H$, it satisfies that
\begin{equation}
{U}^{\otimes(n+1)}\Pi_i (U^\dagger)^{\otimes(n+1)} = \Pi_i,
\end{equation}
for $i=0,\cdots,n$.
\end{Lemma}

{\it Proof.} Suppose POVM $\overrightarrow{\Pi}$ is the optimal
unambiguous programmable discriminator for $n$ states in Hilbert
space $H$. For any unitary matrix $U$ in the Hilbert space $H$,
let $\overrightarrow{\Pi^U}$ be a POVM, such that
\begin{equation}
\Pi^U_i = U^{\otimes(n+1)}\Pi_i (U^\dagger)^{\otimes(n+1)},
\end{equation}
for $i=0,\cdots,n$. Since $\tr_i(\Pi^U_i) = U^{\otimes
n}\tr_i(\Pi_i)(U^\dagger)^{\otimes n} \leq \wedge^n H$,
$\overrightarrow{\Pi^U}$ is clearly also an unambiguous
programmable discriminator. For an arbitrary set of states
$\{\ket{\psi_1},\cdots,\ket{\psi_n}\}$, the success probability of
discriminating them by $\overrightarrow{\Pi}$, is the same as the
success probability of discriminating
$\{U\ket{\psi_1},U\ket{\psi_2},\cdots,U\ket{\psi_n}\}$ by
$\overrightarrow{\Pi^U}$. From Eq.(\ref{def}),
$p(\overrightarrow{\Pi^U}) = p(\overrightarrow{\Pi})$, for any
unitary operator U.

Consider a new measurement $\overrightarrow{\Xi}$, which is the
average of all the above measurements in a unitary distribution
\cite{copy}, i.e.,
\begin{equation}
\Xi_i = \int \mathrm{d}U {U}^{\otimes(n+1)}\Pi_i
(U^\dagger)^{\otimes(n+1)},
\end{equation}
for $i=0,\cdots,n$, where $\mathrm{d}U$ is the normalized positive
invariant measure of the group $U(m)$. Clearly,
$\overrightarrow{\Xi}$ is an unambiguous programmable
discriminator, satisfying that, for any unitary operator $U$ in
$H$, ${U}^{\otimes(n+1)}\Xi_i (U^\dagger)^{\otimes(n+1)} = \Xi_i$,
for $0\leq i\leq n$. Because the efficiency of programmable
discriminators is a concave function,
\begin{equation}
\begin{split}
p(\overrightarrow{\Xi}) &= p\big{(}\int \mathrm{d}U \overrightarrow{\Pi^U} \big{)}\\
&\geq \int \mathrm{d}U p(\overrightarrow{\Pi^U})\\
&= p(\overrightarrow{\Pi}).
\end{split}
\end{equation}
Hence, we can substitute $\overrightarrow{\Pi}$ with
$\overrightarrow{\Xi}$ as the optimal discriminator. \hfill $\Box$

From above lemma, it is known that the optimal unambiguous
programmable discriminators satisfies that
\begin{equation}
U\tr_{\bar{i}}(\Pi_i)U^\dagger = \tr_{\bar{i}}(\Pi_i),
\end{equation}
for any unitary operator $U\in H$. So, $\tr_{\bar{i}}(\Pi_i)$
would be a diagonal matrix.

Next, we provide a relationship between the operators which
consist the measurement for optimal programmable discriminator. In
the total input system of an $n$-state programmable discriminator,
let us denote the $n$ program registers as subsystem $P$, and the
data register as subsystem $D$. Then, we have the following lemma.

\begin{Lemma}\label{LS}
Suppose $\overrightarrow{\Pi}$ is the optimal unambiguous
programmable discriminator for $n$ states, then for any $\sigma\in
S(n)$, it holds that
\begin{equation}
(\sigma^{-1}_P\otimes I_D)\Pi_i(\sigma_P\otimes I_D) =
\Pi_{\sigma_i}
\end{equation}
for $i=1,\cdots,n$.
\end{Lemma}

{\it Proof.} Suppose $\overrightarrow{\Pi}$ is an optimal
unambiguous programmable discriminator. For any $\sigma\in S(n)$,
let $\overrightarrow{\Pi^{\sigma}}$ be a measurement, such that
\begin{equation}
\Pi^{\sigma}_i = (\sigma^{-1}_P\otimes I_D)
\Pi_{\sigma_i}(\sigma_P\otimes I_D),
\end{equation}
for $i\neq 0$. Then, for any $i,j\neq 0$,
\begin{equation}
\begin{split}
&\bra{\psi^n_j}\Pi^{\sigma}_i\ket{\psi^n_j}\\ =
&\bra{\psi_{\sigma_1}}\bra{\psi_{\sigma_2}}\cdots\bra{\psi_{\sigma_n}}\bra{\psi_j}
\Pi_{\sigma_i}\ket{\psi_{\sigma_1}}\ket{\psi_{\sigma_2}}\cdots\ket{\psi_{\sigma_n}}\ket{\psi_j}\\
=&\bra{\tilde{\psi}^n_{\sigma^{-1}(j)}}\Pi_{\sigma_i}\ket{\tilde{\psi}^n_{\sigma^{-1}(j)}},
\end{split}
\end{equation}
where $\ket{\tilde{\psi}_k} = \ket{\psi}_{\sigma_{k}}$, for
$k=1,\cdots,n$. Clearly, $\overrightarrow{\Pi^{\sigma}}$ is also
an unambiguous programmable discriminator, whose efficiency for
discriminating the states $\{\ket{\psi_1},\cdots,\ket{\psi_n}\}$
is equal to the efficiency for discriminating
$\{\ket{\psi_{\sigma_1}},\cdots,\ket{\psi_{\sigma_n}}\}$ by
$\overrightarrow{\Pi}$. Then, the two measurements have the same
efficiency in minimax strategy. Hence, the measurement
$\overrightarrow{\Xi}$, where
\begin{equation}
\Xi_i = \frac{1}{n!}\sum_{\sigma\in S(n)} \Pi^\sigma_i,
\end{equation}
for $1\leq i\leq n$, is an unambiguous programmable discriminator
whose efficiency is no less than $\overrightarrow{\Pi}$. In
addition,
\begin{equation}
(\sigma^{-1}_P\otimes I_D)\Xi_i(\sigma_P\otimes I_D) =
\Xi_{\sigma_i},
\end{equation}
for any $\sigma\in S(n)$. Hence, we can substitute
$\overrightarrow{\Pi}$ by $\overrightarrow{\Xi}$. \hfill $\Box$

From the above two lemmas, it is easy to conclude the following
result.

\begin{Coro}\label{opt}
The optimal unambiguous programmable discriminator
$\overrightarrow{\Pi}$, satisfies that
\begin{equation}
\tr_{\bar{i}}(\Pi_i) = c I_i,
\end{equation}
for $i\neq 0$, where $I_i$ is the identity operator on the $i$th
subsystem, and $c$ is a constant independent of $i$.
\end{Coro}

\section{When the dimension
of state space is equal to the number of discriminated
states}\label{4}

For clarity of presentation, we divide the problem of designing
optimal unambiguous programmable discriminator into two cases. In
this section, we consider the case that the dimension of $H$ is
equal to the number of states to be discriminated. In this
situation, $\wedge^n H$ is a one-dimensional Hilbert space. From
Theorem~\ref{struct}, any unambiguous programmable discriminator
$\overrightarrow{\Pi}$ satisfies that $\Pi_i =
\Pi'_i\otimes\Phi(n)_{\bar{i}}$, where $\Pi'_i$ is a positive
operator on the $i$th subsystem, for any $i\neq 0$. Furthermore,
from Corollary~\ref{opt}, the optimal unambiguous programmable
discriminators satisfies that
\begin{equation}
\Pi_i = c I_i\otimes\Phi(n)_{\bar{i}},
\end{equation}
for $i\neq 0$. Then, we give one of our main results as follows.

\begin{thm}\label{special}
The optimal unambiguous programmable discriminator for $n$ states in
an $n$-dimensional Hilbert space $H$ would be an measurement
$\{\Pi_0,\Pi_1,\ldots,\Pi_n\}$ on the total input space, such that
for $1\leq i\leq n$,
\begin{equation}
\Pi_i = \frac{n}{n+1}I_i\otimes\Phi(n)_{\bar{i}},
\end{equation}
and
\begin{equation}
\Pi_0 = I^{\otimes(n+1)} - \sum_{i=1}^{n}\Pi_i,
\end{equation}
where $I$ is the identity operator on $H$, and $\Phi(n)$ is the
projector of $\wedge^n H$. The success probability of
discriminating states
$\{\ket{\psi_1},\ket{\psi_2},\cdots,\ket{\psi_n}\}$ is
\begin{equation}
p_i =  \frac{n}{(n+1)!}\det(X),
\end{equation}
for any $1\leq i\leq n$, where $X$ is the Gram matrix of states
being discriminated.
\end{thm}

{\it Proof.} Let $\{\ket{1},\ket{2},\ldots,\ket{n}\}$ be an
orthonormal basis of $H$. Then $\Phi(n) = \ket{\phi}\bra{\phi}$,
where
\begin{equation}
\begin{split}
\ket{\phi} &= \ket{1}\wedge\ket{2}\wedge\cdots\wedge\ket{n}\\
&= \frac{1}{\sqrt{n!}}\sum_{\sigma\in S(n)} \sgn(\sigma)
\ket{\sigma_1}\ket{\sigma_2}\cdots\ket{\sigma_n}.
\end{split}
\end{equation}
Consequently,
\begin{equation}
\begin{split}
\Pi_i &= c \sum_{k=1}^n
\ket{k}_i\ket{\phi}_{\bar{i}}\bra{k}_i\bra{\phi}_{\bar{i}}, i \neq 0,\\
\Pi_0 &= I - c\sum_{i=1,k=1}^{n,n}
\ket{k}_i\ket{\phi}_{\bar{i}}\bra{k}_i\bra{\phi}_{\bar{i}}.
\end{split}
\end{equation}

Let $G$ be the Gram matrix of $\{\ket{k}_i\ket{\phi}_{\bar{i}}:1\leq
k\leq n, 1\leq i\leq n\}$, i.e., the $(k,l)$ element in the $(i,j)$
block of matrix $G$ is the inner product of
$\ket{k}_i\ket{\phi}_{\bar{i}}$ and $\ket{l}_j\ket{\phi}_{\bar{j}}$.
When $i=j$, we have
\begin{equation}
\bra{k}_i\bra{\phi}_{\bar{i}}\ket{l}_i\ket{\phi}_{\bar{i}} =
\delta_{k,l},
\end{equation}
and when $i\neq j$, it holds that
\begin{equation}
\begin{split}
\bra{k}_i\bra{\phi}_{\bar{i}}\ket{l}_j\ket{\phi}_{\bar{j}} &=
(-1)^{i-j+1}\frac{(n-1)!}{n!}\delta_{k,l}\\
&= (-1)^{i-j+1}\frac{1}{n}\delta_{k,l}.
\end{split}
\end{equation}
So, the $(i,j)$ block of $G$ is
\begin{equation}
G_{ij} = I\delta_{i,j} + \frac{(-1)^{i-j+1}}{n}I(1-\delta_{i,j}).
\end{equation}

Since the eigenvalues of $\sum_{i=1,k=1}^{n,n}
\ket{k}_i\ket{\phi}_{\bar{i}}\bra{k}_i\bra{\phi}_{\bar{i}}$ are
equal to the eigenvalues of $G$, to confirm $\Pi_0 \geq 0$, the
maximum value of $c$ should be the reciprocal of maximum
eigenvalue of matrix $G$, which can be calculated to be
$\frac{n+1}{n}$ \cite{JF}. As a result, the maximum value of $c$
should be $\frac{n}{n+1}$.

The success probability of discriminating the $i$th state,
\begin{equation}
\begin{split}
p_i &= \bra{\psi^n_i}\Pi_i\ket{\psi^n_i}\\
&= c\bra{\psi'}\Phi \ket{\psi'}\\
&= \frac{c}{n!}\det(X).
\end{split}
\end{equation}
Here
\begin{equation}
\ket{\psi'} =
\ket{\psi_1}\ket{\psi_2}\cdots\ket{\psi_{i-1}}\ket{\psi_{i+1}}\cdots\ket{\psi_n}\ket{\psi_i},
\end{equation}
and $X$ is the Gram matrix of
$\{\ket{\psi_1},\ket{\psi_2},\cdots,\ket{\psi_n}\}$, i.e., the
$(i,j)$ element of $X$,
\begin{equation}
X_{i,j} = \inner{\psi_i}{\psi_j}.
\end{equation}\hfill $\Box$

For any $n$ linearly independent quantum states, let $H$ be the
span space of them, obviously the dimension of $H$ is equal to
$n$. Then, we can design the optimal programmable discriminator
for $n$ states in $H$ by Theorem~\ref{special}, which can
unambiguously discriminate the states. However, it should be noted
that the programmable designed in this way is dependent on the
span space of the states wanted to be discriminated. Although such
a programmable discriminator has a more general utilization than
the discriminator designed according to given states, it also has
an undesirable restriction. An alternative way is to design the
programmable discriminators in a Hilbert space which is so great
that it includes all the states which would be discriminated in
application.

\section{When the dimension
of state space is greater than the number of discriminated
states}\label{5}

In this section, we consider the problem of designing unambiguous
programmable discriminators for $n$ states in an $m$-dimensional
Hilbert space $H$, where $m > n$. In this case, the structure of
optimal unambiguous programmable discriminators is not clear by now.
We conjecture that they have a similar structure to that of optimal
discriminators for the case that $m=n$, i.e.,
\begin{equation}\label{hp}
\Pi_i = cI_i\otimes\Phi(n)_{\bar{i}},
\end{equation}
for $i\neq 0$. Clearly, this structure satisfies the demands
offered by Lemma~\ref{LU} and Lemma~\ref{LS}. The remainder of
this section is devoted to give the optimal one of discriminators
satisfying Eq.(\ref{hp}).

Suppose $\{\ket{1},\ket{2},\ldots,\ket{m}\}$ is an orthonormal basis
for Hilbert space $H$. Let $\Sigma_n$ denote the set of all strictly
increasing $n$-tuples chosen from $\{1,2,\ldots,m\}$, i.e.,
$\varsigma=(\varsigma_1,\varsigma_2,\ldots,\varsigma_n)\in\Sigma_n$
if and only if $1\leq \varsigma_1 < \varsigma_2 < \cdots <
\varsigma_n \leq m$. For all $\varsigma\in\Sigma_n$, let
\begin{equation}
\begin{split}
\ket{\phi_\varsigma} &= \ket{\varsigma_1} \wedge \ket{\varsigma_2}
\wedge \cdots \wedge \ket{\varsigma_n}\\
&= \sum_{\sigma\in S(n)}\sgn(\sigma)
\ket{\varsigma_{\sigma_1}}\ket{\varsigma_{\sigma_2}}\cdots\ket{\varsigma_{\sigma_n}}
\end{split}
\end{equation}
$\ket{\phi_\varsigma}$s construct an orthonormal basis for
$\wedge^n H$, i.e., $\Phi(n) = \sum_{\varsigma\in\Sigma_n}
\ket{\phi_\varsigma}\bra{\phi_\varsigma}$, and
\begin{equation}
\Pi_i = c\sum_{1\leq k\leq m,\varsigma\in\Sigma_n}
\ket{k}_i\ket{\varsigma}_{\bar{i}}\bra{k}_i\bra{\varsigma}_{\bar{i}},
\end{equation}
for $i \neq 0$.

Analogous to the situation that $m=n$, the maximum value of $c$ is
the reciprocal of maximum eigenvalue of the Gram matrix of
$\{\ket{k}_i\ket{\varsigma}_{\bar{i}}\}$, where $1\leq i \leq n$,
$1\leq k \leq m$, and $\varsigma\in\Sigma_n$. The elements of this
Gram matrix can be expressed as
$\bra{k}_i\bra{\phi_\varsigma}_{\bar{i}}\ket{l}_j\ket{\phi_\tau}_{\bar{j}}$.

First, if $i=j$,
\begin{equation}
\bra{k}_i\bra{\phi_\varsigma}_{\bar{i}}\ket{l}_i\ket{\phi_\tau}_{\bar{i}}
= \delta_{k,l}\delta_{\varsigma,\tau}.
\end{equation}

Next, if $i\neq j$, $\varsigma = \tau$, and $k,l\in\varsigma$,
\begin{equation}
\bra{k}_i\bra{\phi_\varsigma}_{\bar{i}}\ket{l}_j\ket{\phi_\varsigma}_{\bar{j}}
= (-1)^{i-j+1} \frac{1}{n}\delta_{k,l}.
\end{equation}

In addition, if the condition $i\neq j$ also holds, and there
exists $\xi\in\Sigma_{n+1}$, i.e, $\xi$ is an $(n+1)$-tuple chosen
from $\{1,2,\ldots,m\}$, satisfying that $\xi = \{k\}\cup\varsigma
= \{l\}\cup\tau$,
\begin{equation}
\bra{k}_i\bra{\phi_\varsigma}_{\bar{i}}\ket{l}_j\ket{\phi_\tau}_{\bar{j}}
= (-1)^{j-i+\xi^{-1}(k)-\xi^{-1}(l)}\frac{1}{n}(1-\delta_{k,l}),
\end{equation}
where $\xi^{-1}(k)$, $\xi^{-1}(l)$ denote the position of $k$, $l$
in the strict increasing $(n+1)$-tuple $\xi$, respectively.

Finally, all other elements
$\bra{k}_i\bra{\phi_\varsigma}_{\bar{i}}\ket{l}_j\ket{\phi_\tau}_{\bar{j}}$
in this matrix would be zero.

Therefore, the Gram matrix is
\begin{equation}
G = (\bigoplus_{\varsigma}\Gamma_\varsigma)\bigoplus
(\bigoplus_{\xi} \Lambda_\xi).
\end{equation}
Here $\Gamma_\varsigma$ is the Gram matrix of
$\{\ket{k}_i\ket{\phi_\varsigma}_{\bar{i}}\}$, where
$k\in\varsigma$, $\varsigma\in\Sigma_n$; $\Lambda_\xi$ is the Gram
matrix of $\{\ket{k}_i\ket{\phi_{\xi-\{k\}}}_{\bar{i}}\}$, where
$k\in\xi$, $\xi\in\Sigma_{n+1}$, and $\xi-\{k\}$ denotes the
strictly increasing $n$-tuple comprised of the elements in $\xi$
except $k$. The maximum eigenvalue of $G$ is the greatest one of
eigenvalues of $\Gamma_\varsigma$s and $\Lambda_\xi$s.

The $(i,j)$ block of matrix $\Gamma_\varsigma$ is
\begin{equation}
I\delta_{i,j} + \frac{(-1)^{i-j+1}}{n}I(1-\delta_{i,j}),
\end{equation}
and the maximum eigenvalue of $\Gamma_\varsigma$ is
$\frac{n+1}{n}$.

The $(k,l)$ element of the $(i,j)$ block in matrix $\Lambda_{\xi}$
is
\begin{equation}
\delta_{i,j}\delta_{k,l}
+\frac{(-1)^{i-j+k-l}}{n}(1-\delta_{k,l})(1-\delta_{i,j}),
\end{equation}
and the maximum eigenvalue of $\Lambda_\xi$ can be calculated to
be $n$.

Consequently, the maximum value of $c$ should be $\frac{1}{n}$.
Hence, the optimal one of unambiguous programmable discriminators
for $n$ quantum states in a $m$-dimensional Hilbert space $H$, which
has the form given in Eq.(\ref{hp}), is a measurement
$\{\Pi_0,\Pi_1,\ldots,\Pi_n\}$ on the total input system, such that
for $1\leq i\leq n$,
\begin{equation}\label{gnz}
\Pi_i = \frac{1}{n}I_i\otimes\Phi(n)_{\bar{i}},
\end{equation}
and
\begin{equation}\label{gz}
\Pi_0 = I^{\otimes(n+1)} - \sum_{i=1}^{n}\Pi_i,
\end{equation}
where $I$ is the identity operator on $H$, and $\Phi(n)$ is the
projector on $\wedge^n H$. Moreover, the success probability of
discriminating states
$\{\ket{\psi_1},\ket{\psi_2},\cdots,\ket{\psi_n}\}$ is
\begin{equation}
\begin{split}
p = \frac{1}{n \cdot n!} \det(X),
\end{split}
\end{equation}
where $X$ is the Gram matrix of states being discriminated.

It is easy to see that the success probability of discriminating a
set of states is not related to the dimension of $H$, so we can
choose $H$ to be a great enough Hilbert space in order to include
all quantum states which may occur in application. Then, the
unambiguous programmable discriminator given by Eq.(\ref{gnz}) and
Eq.(\ref{gz}) is suitable for any $n$ states under consideration.

The success probability of this discriminator turns out to be
zero, if and only if the states to be discriminated are linearly
dependent. As we know, the necessary and sufficiency condition for
a set of states to be unambiguously discriminated is that the
states are linearly independent \cite{CH}. So, the states which
cannot be unambiguously discriminated by our devices are also
unable to be unambiguously discriminated by any other device. In
this way, we can claim that our programmable discriminators are
universal.

On the other hand, in the minimax strategy, if we know the exactly
set of states being discriminated, the optimal success probability
for unambiguously discriminating $n$ states $\{\ket{\psi_i}\}$ is
the minimum eigenvalue of $X$, where $X$ is the Gram matrix of
$\{\ket{\psi_i}\}$\cite{SZ,Minimax}. Let $p_s$ denote this optimal
efficiency, and $p$ denote the efficiency of the universal
unambiguous programmable discriminator for the same states.
Because $(p_s)^n \leq \det(X) \leq p_s$, , it holds that
\begin{equation}
\frac{1}{n\cdot n!} (p_s)^n \leq p \leq \frac{1}{n\cdot n!} p_s.
\end{equation}
Hence, when $n$ is large, the efficiency of programmable
discriminator would be quite undesirable, comparing to the
discriminator especially designed to known states.

\section{An application to mixed states}\label{6}
In this section, we will show how to use the discriminators given
above to unambiguously discriminate a set of mixed states.
Different from the discrimination for pure states, when the states
to be discriminated are mixed, we have to prepare extra states for
quantum program.

First, we would like to introduce the notion of ``core''s for a
set of mixed states, which first be mentioned in Ref.
\cite{mymixed}. For $n$ mixed states $\rho_1,\cdots,\rho_n$, their
``core''s are defined as follows. From Ref. \cite{mymixed}, any
states $\rho_i$ can be uniquely divided into two parts,
\begin{equation}
\rho_i = \tilde{\rho}_i + \hat{\rho}_i,
\end{equation}
such that
\begin{equation}
\fsup(\hat{\rho}_i) \leq \sum_{j\neq i}\fsup(\rho_j),
\end{equation}
and
\begin{equation}
\fsup(\tilde{\rho}_i) \cap \sum_{j\neq i}\fsup(\rho_j) = 0.
\end{equation}
Consequently, let $\tilde{\rho}_0=\sum_{i=1}^n \hat{\rho}_i$.
Then, we call
$\tilde{\rho}_0,\tilde{\rho}_1,\cdots,\tilde{\rho}_n$ the
``core''s of states $\rho_1,\cdots,\rho_n$.

The quantum program we used is comprised of $n+1$ linearly
independent state sets, $S_0,S_1,\cdots,S_n$, satisfying that
$S_i$ can give rise to $\tilde{\rho}_i$ with a corresponding
probability distribution, for $i=0,\cdots,n$, i.e.,
$S_i=\{\ket{\psi^i_1},\cdots,\ket{\psi^i_{m_i}}\}$, where the
states in $S_i$ are linearly independent, and
\begin{equation}
\tilde{\rho}_i = \sum_{j=1}^{m_i}
q_{ij}\ket{\psi^i_j}\bra{\psi^i_j},
\end{equation}
with some coefficients $\{q_{ij}\}$ satisfying that
$\sum_{j=1}^{m_i} q_{ij} = \tr(\tilde{\rho}_i)$. Obviously, $m_i$
is the dimension of support space of $\tilde{\rho}_i$. The program
can be denoted by
\begin{equation}
\ket{\Psi} =
\ket{\psi^0_1}\cdots\ket{\psi^0_{m_0}}\cdots\ket{\psi^n_1}\cdots\ket{\psi^n_{m_n}}.
\end{equation}

Let $N = \sum_{i=0}^n m_i$, then the $n$ mixed states can be
unambiguously discriminated by an unambiguous programmable
discriminator for $N$ pure states. In the discriminator, we first
partition the $N$ program registers into $n+1$ parts, labeled from
$0$ to $n$. The $j$th register in the $i$th part is put in the
state $\ket{\psi^i_j}$. When the state in data register is
$\rho_s$, then the possibility of having outcome in the $i$th part
is
\begin{equation}
\begin{split}
p_i &= \sum_j \tr(\Pi^i_j \ket{\Psi}\bra{\Psi}\otimes\rho_s)\\
&= \sum_j
\tr(\Pi^i_j\ket{\Psi}\bra{\Psi}\otimes\tilde{\rho_s})+\tr(\Pi^i_j\ket{\Psi}\bra{\Psi}\otimes\hat{\rho_s}).
\end{split}
\end{equation}
Because $0\leq\hat{\rho_s}\leq\tilde{\rho}_0$, there is an upper
bound and a lower bound of $p_i$. On the one hand,
\begin{equation}\label{m1}
\begin{split}
p_i &\geq \sum_j \tr(\Pi^i_j\tilde{\rho_s})\\
&= \sum_j\sum_k q_{sk}
\bra{\Psi}\bra{\psi^s_k}\Pi^i_j\ket{\Psi}\ket{\psi^s_k}\\
&= \sum_j\sum_k q_{sk}\frac{1}{N\cdot N!}\det(X)\delta_{is}\delta{jk}\\
&= \frac{\tr(\tilde{\rho}_s)}{N\cdot N!}\det(X)\delta_{is},
\end{split}
\end{equation}
where $X$ is the Gram matrix of the states in $\ket{\Psi}$.
$\det(X)$ is always greater than zero, since the states in
$\ket{\Psi}$ is designed to be linearly independent. On the other
hand,
\begin{equation}
\begin{split}
p_i &\leq \sum_j \tr(\Pi^i_j \tilde{\rho_s})+ \tr(\Pi^i_j \tilde{\rho_0})\\
&= \frac{\tr(\tilde{\rho}_s)}{N\cdot N!}\det(X)\delta_{is} +
\frac{\tr(\tilde{\rho}_0)}{N\cdot N!}\det(X)\delta_{i0}.
\end{split}
\end{equation}
Hence, when the state being identified is $\rho_s$, the
measurement result can only happen in the $s$ part or the $0$
part. If we consider the latter situation as an inconclusive
answer, then this scheme is a well-defined unambiguous
discrimination for the mixed states. Moreover, from
Ref.{\cite{mymixed}}, a sufficient and necessary condition for
unambiguously discriminating the mixed states
$\{\rho_1,\cdots,\rho_n\}$ with a non-zero success probability is
that for any $i\neq 0$, $\tilde{\rho}_i\neq 0$, which is
equivalent to that the probability of getting a result in the $s$
part, in other words the success probability of our scheme, is
always greater than 0.

\section{Summary}\label{7}

In this paper, the problem of designing programmable
discriminators for any $n$ quantum states in a given Hilbert space
$H$ is addressed. First, we give a necessary and sufficient
condition for judging whether a measurement is an unambiguous
programmable discriminator. Then, by utilizing the minimax
strategy to evaluate the efficiency of discrimination, we offer
several conditions for the optimal programmable discriminators,
and give the optimal programmable discriminator in the case that
the span space of the states is known. Furthermore, we propose a
universal programmable discriminator, which can unambiguously any
$n$ states under consideration. We also give another application
of these universal program discriminators: they can be used to
unambiguously discriminate a set of mixed states.

\end{document}